\documentclass{ws-ijmpa}

\begin{document}

\markboth{T.~Han, G.~Valencia and Yili Wang}
{New Signatures For Top In Hadron Collider}

\catchline{}{}{}{}{}
%
%%%%%%%%%%%%%%%%%%%%%%%%%%%%%%%%%%%%%%%%%%%%%%%%%%%%%%%%%%%%%%%%%%%%

\title{New Signatures For Top In Hadron Collider
}

\author{\footnotesize Tao Han \footnote{than@pheno.physics.wisc.edu}}

\address{Department of Physics, University of Wisconsin, Madison,
WI  53706,\\ and 
Institute of Theoretical Physics, Academia Sinica, Beijing 100080, China
}

\author{\footnotesize German Valencia \footnote{valencia@iastate.edu} and \footnotesize Yili Wang \footnote{yiliwa@iastate.edu, conference speaker}}

\address{Department of Physics and Astronomy, Iowa State University \\
Ames, Iowa 50010
}
\maketitle
\pub{Received (27 10 2004)}{}
\begin{abstract}
We study the signatures for new TeV resonances that couple to top or bottom 
quarks both at the Tevatron Run II and at the LHC. We find that it is possible 
to study these resonances when they are produced in association with a pair of 
heavy quarks or in association with a single top at the LHC. In particular, with an integrated luminosity of 300 fb$^{-1}$ at the LHC, it is possible to probe resonance masses up to around 2~TeV.

\keywords{Top Physics; Electroweak Interaction; Beyond SM.}
\end{abstract}
~\linebreak
If there is no light Higgs boson found in the next generation of collider experiments, the interactions among the longitudinal vector bosons would become strong at a scale of ${\cal O}$(1 TeV)
and new dynamics must set in. The fact that the top-quark
mass is very close to the electroweak scale 
($m_t \approx v/\sqrt 2$) is rather suggestive: there may be a 
common origin for electroweak symmetry
breaking and top-quark mass generation.
A major goal for the Fermilab Tevatron and the CERN LHC is the detailed study
of the properties of the top quark. In particular they should
establish whether the third family behaves like the first two, or whether it is subject to new interactions.

This paper is a brief summary of a study reported elsewhere \cite{Han:2004zh}, in which we introduce  new resonances which couple strongly to the third generation  but not necessarily to the $W$ and $Z$ gauge bosons. We are interested in new interactions 
of $b$ and $t$ quarks with these new heavy resonances. Our goal in this paper is to investigate the extent to which hadron colliders are sensitive to new interactions of the top quark regardless of the origin of the 
new interactions.
 
We begin with new vector resonance. We assume that this vector resonance has negligible couplings to electroweak gauge bosons. 
 We thus proceed with the following effective Lagrangian 
coupling a spin one field to the top and bottom quarks,
\begin{equation}
{\cal L} = - \Psi \gamma^\mu (g_V + g_A \gamma_5)\tau_i \Psi V^i_\mu.
\label{simplevp}
\end{equation}

In hadron colliders, however, light quark annihilation represents a 
significant production source for new vector resonances even if they 
couple predominantly to $b$ and $t$. To keep our study as model-independent as possible we choose a frame that make the contributions of the light quark annihilation  small. 
For definiteness we will use the $Z^\prime$ model of 
Ref.~\cite{zprime} in the limit of no $V-Z$ mixing. 
\begin{eqnarray} 
{\cal L} &=& 
{g\over 2}\tan\theta_W \tan\theta_R  
({1\over 3} \bar q_L \gamma^\mu q_L+ {4\over 3} \bar u_{Ri} \gamma^\mu u_{Ri}
-{2\over 3} \bar d_{Ri}\gamma^\mu d_{Ri}) V_\mu \nonumber\\
&-& {g\over 2}\tan\theta_W  (\tan\theta_R + \cot\theta_R) (
\bar t_{R} \gamma^\mu t_{R} - 
\bar b_{R} \gamma^\mu b_{R}) V_\mu . 
\label{neucoupl}
\end{eqnarray}
In this expression $\theta_R$ is a new parameter. 
With large $\cot\theta_R$, this model provides a specific 
example of a new vector resonance which
couples to $b$ and $t$ significantly and couples to the light fermions weakly. 
 In the limit of large $\cot\theta_R$ 
these couplings are purely right-handed, with 
$g_A = g_V = (g / 4)\tan\theta_W\cot\theta_R$, in which 
 $\cot\theta_R$ must be smaller than 20 if the new interaction remains perturbative.

\begin{figure}[ht]
\centerline{\psfig{file=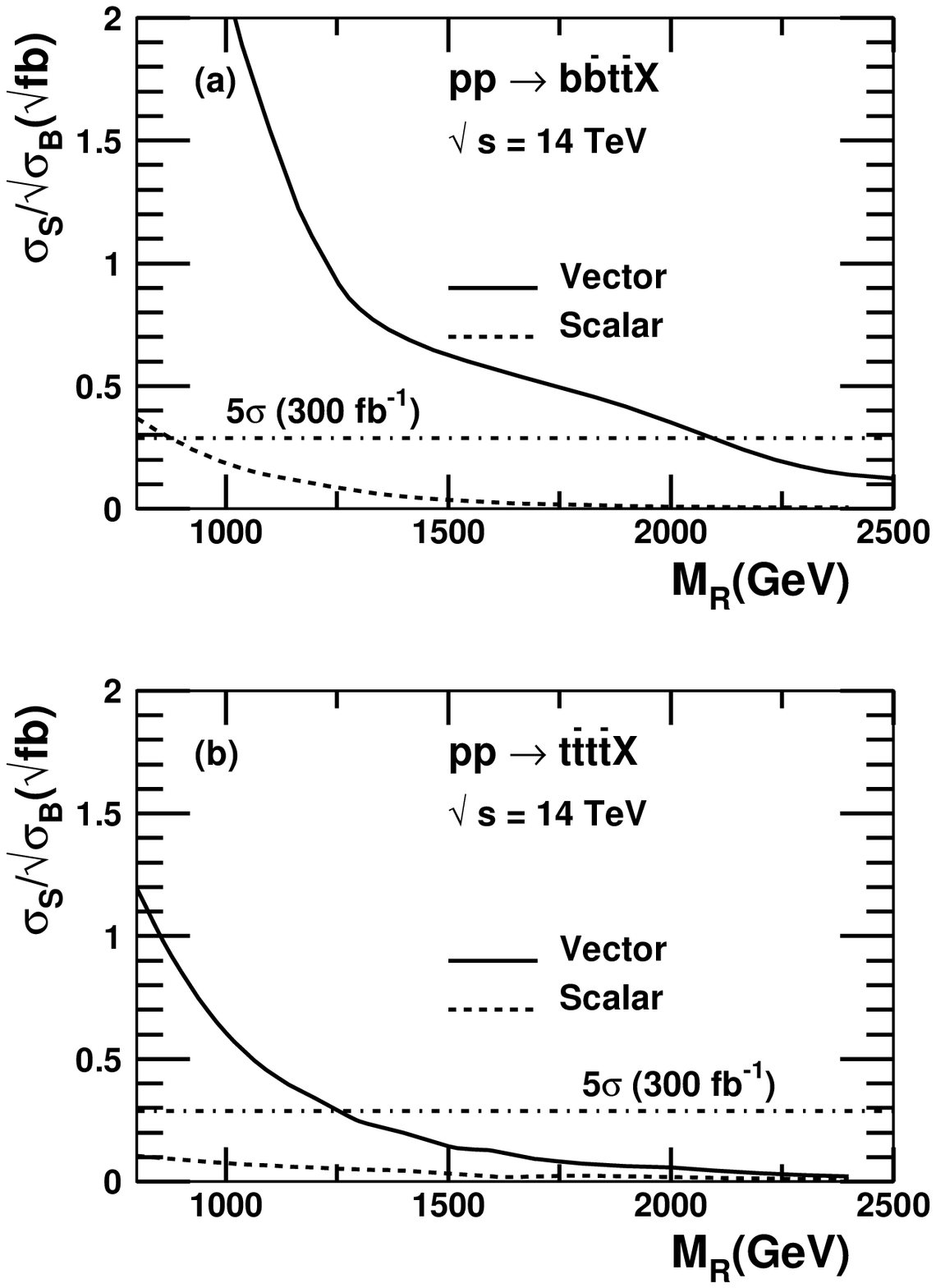,width=5.cm} \hspace{1.5cm}\psfig{file=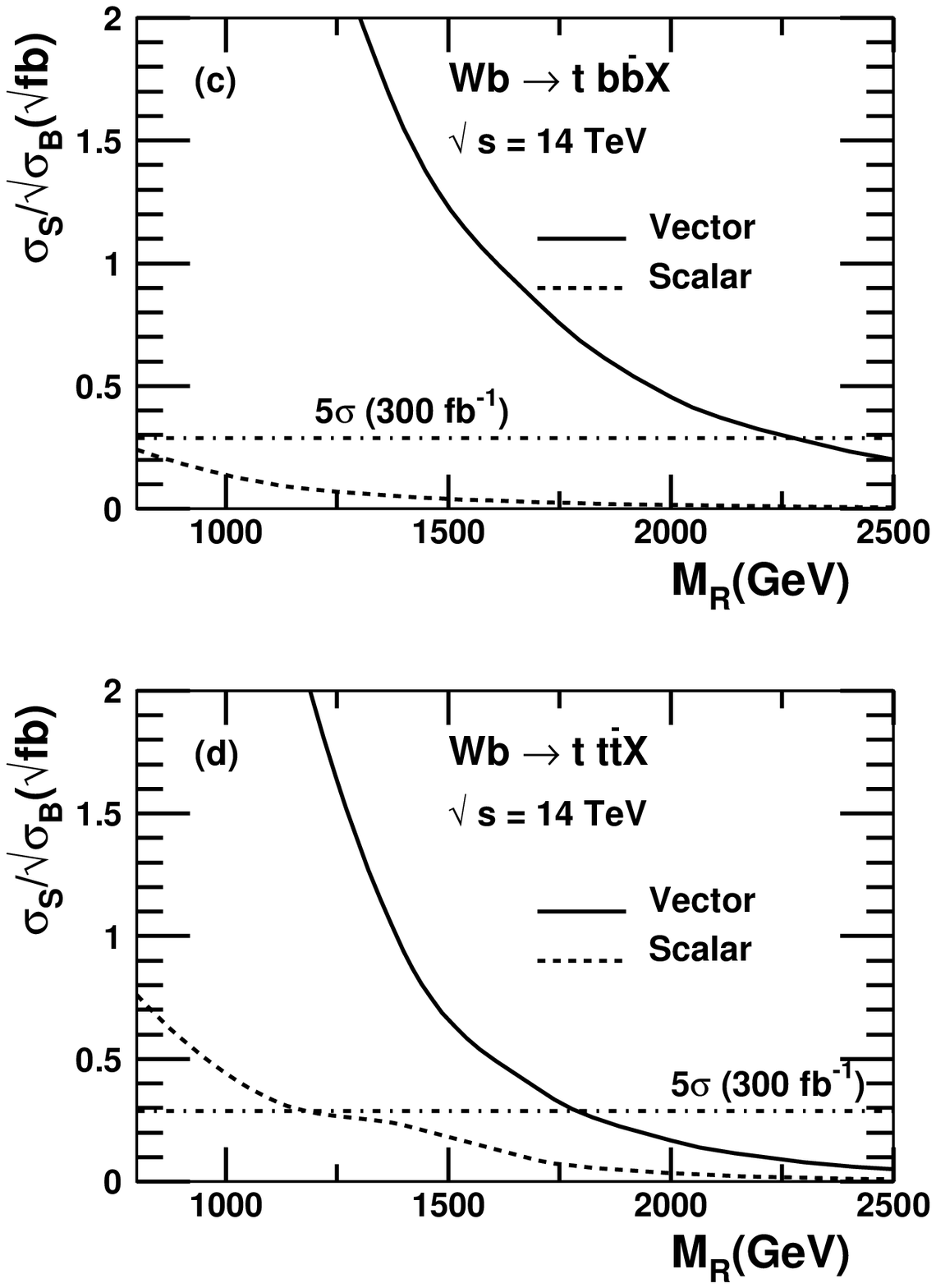,width=5.cm}}
\caption{ 
Statistical sensitivity at the LHC with  an integrated luminosity of 300 fb$^{-1}$ 
to a new resonance as a function of the 
resonance mass for (a) $pp \rightarrow b \bar{b} t\bar{t}X$, (b)  $pp \rightarrow t \bar{t} t\bar{t}X$, (c) $w \rightarrow t, b\bar{b}$, and (d) $W \rightarrow t, t\bar{t} $.} 
\label{fig:reach}
\end{figure}

We next consider  a new scalar resonance. In this case we use 
a very simple (non-renormalizable) parametrization for the new interactions,  
and assume that the couplings of the scalar to the light fermions are 
completely negligible. We write 
\begin{equation}
{\cal L}= - {m_t\over v} S \left( \kappa_b \bar b b + \kappa_t \bar t t\right).
\label{lstt}
\end{equation}
This form allows us to parameterize simultaneously the cases 
where either the $b$-quark or the $t$-quark or both have enhanced 
couplings to the new scalar resonance.
The tightest constraint is obtained by requiring perturbative 
unitarity~\cite{han} in the scattering amplitude 
$b\bar{b} \rightarrow b \bar{b}$ 
(or in $t\bar{t} \rightarrow t \bar{t}$) 
through an exchange of the new scalar. This leads to 
$\kappa_{b,t} \leq 3$.

we compute the contributions of the new interactions 
to multiple processes and conclude that Tevatron is not sensitive to this type of new physics. We find that processes with four heavy quarks, originating in the production of the heavy resonance in association with two heavy quarks are the most promising channels for the signal searchs in LHC.
In the 4-quark process $pp \rightarrow b\bar{b} t \bar{t} X$ or $t\bar{t} t \bar{t} X $, the signal is completely dominated by the gluon fusion. There is also a   much smaller contribution initiated by $b\bar b$ annihilation that we have calculated but not included. 
We use COMPHEP~\cite{comphep} to compute the signal and MADGRAPH~\cite{Stelzer:1994ta} to compute the corresponding SM background. We implement cuts 
\begin{eqnarray}
p_T(b) > 100 {\rm ~GeV}, \quad p_T(t) > 50 {\rm ~GeV}, \quad  |y_b|  <  2, \quad M_R - 4\Gamma_{R}  <  m_{b\bar{b}, t\bar{t}} <  M_R + 4\Gamma_{R} \nonumber 
\label{bbtt-cuts}
\end{eqnarray}
We present the sensitivity at the LHC assuming a total integrated luminosity of  300 fb$^{-1}$ in Fig.~\ref{fig:reach}.  With $50\%$ of $b\bar{b}$-tagging efficiency and $0.5\%$ faked b rate~\cite{rate}, we end up a combined event efficiency of about $16\%$ for $t\bar{t}$. 
We see that a $5\sigma$ sensitivity may be reached for masses up to 2~TeV 
for the vector resonance. 

It is well known that single top quark production via the 
electroweak process $Wb \rightarrow t$ can be sizable 
due to the enhanced longitudinal gauge boson coupling at high energies. 
The cross section for single top production increases with energy 
up to about one-third of the cross section for 
$t\bar{t}$ pair production \cite{sintop}. The main 
advantage of this channel is the substantially smaller standard model 
background.
We consider the $Wb \to t$ process, associated with a heavy resonance that radiates off the top quark and decays to $b \bar{b}$ or $t\bar{t}$.  We use cuts: 
$p_T(t, b) > 100$~GeV and $|y_{t,b}|  < 2 $. 
The high $p_T$ cut is imposed  on {\it all} heavy quarks,
including the two b quarks that reconstruct the resonance mass as well as the single top quark. Using the same $b$ and $t$ efficiencies as before we show
in Fig.~\ref{fig:reach} the sensitivity in this case. The reach can be up to $M_V \sim 2$~TeV with a $5\sigma$ significance.

\section*{Acknowledgments}
The work of T.H. was supported in part 
by the US DOE under contract No. DE-FG02-95ER40896, in part by the
Wisconsin Alumni Research Foundation, and in part by
National Natural Science Foundation of China.
The work of G.V. and Y. W. was supported
in part by DOE under contact number DE-FG02-01ER41155.

\end{document}